\documentclass{aa}  

\usepackage{soul}
\usepackage{pdfpages}
\usepackage{graphicx}
\usepackage{epstopdf}
\usepackage{txfonts}
\begin{document}

   \title{Hydrodynamical instabilities induced by atomic diffusion in A stars and their consequences}

   \subtitle{}

   \author{ Morgan Deal\inst{1,3}, Olivier Richard\inst{1}, \and Sylvie Vauclair\inst{2,3}}
\institute{Laboratoire Univers et Particules de Montpellier (LUPM), UMR 5299, Universit\'e de Montpellier, CNRS, Place Eug\`ene Bataillon, 34095 Montpellier Cedex 5, France\     
           \and
           Universit\'e de Toulouse, UPS-OMP, IRAP, France
           \and
           CNRS, IRAP, 14 avenue Edouard Belin, 31400 Toulouse, France\\
            \email{morgan.deal@umontpellier.fr} 
           }

   \date{}

 
  \abstract
   {}
   {Atomic diffusion, including the effect of radiative accelerations on individual elements, leads to important variations of the chemical composition inside the stars. The accumulation in specific layers of the elements, which are the main contributors of the local opacity, leads to hydrodynamical instabilities that modify the internal stellar structure and surface abundances. Our aim is to study these effects and compare the resulting surface abundances with spectroscopic observations }
   {We computed the detailed structure of A-type stars including these effects. We used the Toulouse-Geneva Evolution Code (TGEC), where radiative accelerations are computed using the Single Valued Parameter (SVP) method, and we added double-diffusive convection with mixing coefficients deduced from three-dimensional (3D) simulations.}
   {We show that the modification of the initial chemical composition has important effects on the internal stellar mixing and leads to different surface abundances of the elements. The results fit the observed surface chemical composition well if the layers, which are individually mixed by double-diffusive convection, are connected.}
   {}

   \titlerunning{hydrodynamical instabilities induced by atomic diffusion}
  
   \authorrunning{Deal et al.}  

   \maketitle
%

\section{Introduction}

Atomic diffusion in stars is a fundamental process that has been discussed by the pioneers of stellar physics \citep{eddington26}. In the standard computations of stellar internal structure, the basic equations are written as if the stars were composed of only one chemical element representing the global mixture. An average atomic mass, namely the mean molecular weight $\mu$, is introduced in the equation of hydrostatic equilibrium. The equation of radiative transfer is also computed for a unique average chemical element with a unique absorption coefficient, i.e. the Rosseland mean opacity, representing the average effect of all the ions present in the stellar plasma. 

Real stars are composed of many different chemical species. Each of them behaves its own way, according to its own molecular weight, the electric charge, and the way it absorbs photons. This behaviour changes with depth, according to the ionisation stage of the element. Meanwhile, all the elements feel the same global pressure and temperature gradients, which adjust according to the average parameters of the global gas. 

As a result of the selective effects acting on individual ions, these ions selectively move along a small distance, during a small timescale, before sharing the acquired momentum to their surroundings through collisions. Atomic diffusion is the statistical result of a large number of these individual motions. This is ignored in standard models that only take what happens after the momentum is collectively shared into account. 

In the 1970's, atomic diffusion was suggested to be responsible for a large variety of chemical peculiarities observed in stars \citep{michaud76,vauclair82,michaud83, richer00,michaud08}. Later on, it was shown that the importance of atomic diffusion is not restricted to the so-called "peculiar stars" and that this process occurs in all kinds of stars, although it is moderated by other competing transport processes. A spectacular confirmation of the importance of atomic diffusion inside the Sun was given by helioseismology \citep{bahcall95,richard96,gough96}. This process was also proved to have a major impact on the pulsational properties of various types of stars \citep{charpinet97,turcotte00,alecian09}. 

At the present time, most stellar evolution codes introduce helium gravitational settling in their computations. Some of these codes also take the atomic diffusion of heavy elements into account in a simplified way without computing the effects of radiative accelerations. Very few codes include the computations of the selective atomic diffusion of many individual species in a complete way, taking into account all the competing effects such as gravitational, thermal and radiative effects, and those effects due to the concentration gradients \citep{turcotte98,hui-Bon-Hoa08,theado12}. 

First of all, these accumulation and depletion processes strongly influence the local opacities. The local abundance increase of iron and nickel, which represent important contributors to the opacity in some stellar layers, may lead to extra convective zones \citep{richer00,richard01} and can even, in some cases, trigger stellar pulsations through the iron-induced $\kappa$-mechanism \citep{charpinet97,pamyatnykh04,bourge06,alecian09}.  

The iron and nickel  accumulations that are radiatively induced precisely occur at the place of the so-called "opacity bump" because the local variations of the radiative accelerations are directly related to the local opacity increase. As a consequence, the induced accumulation leads to a still larger local opacity. 

The modifications of the abundances of individual elements due to atomic diffusion also lead to variations of the local mean molecular weight. As helium is never supported by the radiative flux in its original abundance, it always settles down, which introduces a stabilizing contribution to the local $\mu$-gradient. On the other hand, heavy element accumulation leads to inverse, destabilizing $\mu$-gradients. When the global mean molecular weight increases towards the surface, the local stellar gas is subject to thermohaline, or fingering convection \citep[ and references therein]{vauclair04, garaud11, deal13}.

The situation may be summarised as follows: the radiative accelerations acting on individual elements push them upwards so that they selectively travel during a short timescale before sharing the acquired momentum with the surroundings. The resulting radiative effect on the global gas is small. However the accumulation of heavy elements induced by the selective process leads to an increased mean molecular weight, which can lead to macroscopic mixing through fingering convection. 

Spectroscopic observations indicate that the efficiency of atomic diffusion is generally reduced in the envelope of stars by competing transport processes. Several macroscopic processes have been tested, including rotational instabilities of various kinds, internal gravity waves, and mass loss. \citep{talon06,vick10}.
In all these studies, fingering convection was bypassed. This process should be the first  tested in the computations, as, contrary to the other processes, it is a direct consequence of atomic diffusion. 

In the present paper, we study, for three A-star models of different masses, the consequences of introducing the effect of fingering convection and we compare the results to abundance determinations obtained from observations of few A stars.

\section{Numerical computations}
\subsection{Stellar models}

The stellar models were computed using the Toulouse-Geneva Evolution Code (TGEC), which includes atomic diffusion with radiative accelerations calculated for 21 species, namely 12 elements and their main isotopes: H, $^{3}$He, $^{4}$He, $^{6}$Li, $^{7}$Li, $^{9}$Be, $^{10}$B, $^{12}$C, $^{13}$C, $^{14}$N, $^{15}$N, $^{16}$O, $^{17}$O,$^{18}$O, $^{20}$Ne, $^{22}$Ne, $^{24}$Mg, $^{25}$Mg, $^{26}$Mg, $^{40}$Ca, and $^{56}$Fe \citep{theado12}. The diffusion computations are based on the Boltzmann equation for a dilute collision-dominated plasma. When the medium is isotropic, the solution of the Boltzmann equation is a Maxwellian distribution function. In stars however, structural gradients (temperature, pressure, density, etc.) lead to small deviations from the Maxwellian distribution, which are specific to each species. Solutions of the Boltzmann equation are then obtained in terms of convergent series representing successive approximations to the true distribution function \citep{chapman70}. The computations lead to a statistical "diffusion" or "drift" velocity $w_d$ of the element with respect to the main component of the plasma. The \cite{montmerle76} treatment is used for helium. The abundance variations of all the elements are computed simultaneously with the use of the mass conservation equation. The gravitational and thermal diffusion coefficients used in the code are those derived by \citet{paquette86}. The equation of state used in the code is the OPAL2001 equation \citep{rogers02}. The nuclear reaction rates are from the NACRE compilation \citep{angulo99}. 

\subsubsection{Opacities and radiative accelerations}

We use OPCD v3.3 codes and data \citep{seaton05} to compute Rosseland opacity at each time step to take the variations of the abundances of each element into account. We use the opacity data to compute individual radiative accelerations on C, N, O, Ne, Mg, Ca, and Fe. This is carried out with the improved semi-analytical prescription proposed by \citet{alecian04}. Radiative accelerations due to bound-bound (\citealt{alecian85}; \citealt{alecian90}) and bound-free \citep{alecian94} transitions are obtained using a parametric form of the radiative acceleration equation. The basic idea of this parametric method is to derive formula in which the terms depending explicitly on atomic data (such as \textit{gf} values) are separated from those depending on the stellar plasma and abundances of the considered ion with the aim of accounting for saturation effects. In this framework, the radiative accelerations may be approximated by calculating a single value for each parameter found in the related equations. This is the so-called Single Valued Parameter (SVP) approximation \citep{leblanc04}. The interface of the SVP method with the models consists of a set of six parameters per ion, which allows us to estimate the radiative acceleration of each element through simple algebraic expressions \citep{alecian04} for each time step of the run of the evolution code. These parameters are determined at the beginning of the computation through interpolation as a function of the stellar mass inside a pre-established grid. The computation of the total radiative acceleration for a given element with SVP also requires computing the relative populations of the ions. This is included in the set of the SVP numerical routines added to TGEC. 

\subsubsection{Convection and mixing}

In our models, dynamical convection zones are computed using the mixing length formalism with a mixing length parameter of 1.8. There are assumed to be instantaneously homogenised. The HI and HeII convective zones are supposed to be connected by overshooting and mixed together \citep{latour81}. The iron convective zone, appearing in some models, is supposed to be disconnected from the surface convective zone.
The models are evolved from pre-main sequence up to hydrogen core exhaustion, and atomic diffusion is introduced at the beginning of the main sequence.

To avoid the appearance of steep and unrealistic abundance gradients at the transition between radiative and convective regions, we introduce mild mixing at the bottom of each convective zone. We use a mixing diffusion coefficient of the form \citep{theado09}
\begin{equation}
D_{mix}=D_{bcz}~\rm{exp} \left(\frac{r-r_{bzc}}{\Delta} \ln2\right)
,\end{equation}

where $D_{bcz}$ and $r_{bcz}$ are respectively the value of $D_{mix}$ and the value of the radius at the bottom of the convective zone. The value of $D_{bcz}$ is taken as $10^5$ $\rm{cm}^2~\rm{s}^{-1}$ . That of $\Delta$ is taken as 0.2 \% of the stellar radius below dynamical convective zones and to 0.05 \% of the stellar radius below fingering convective zones. These values are chosen to avoid discontinuity in the effective turbulent diffusion coefficient.

\begin{figure*}
\begin{center}
\includegraphics[width=0.95\textwidth]{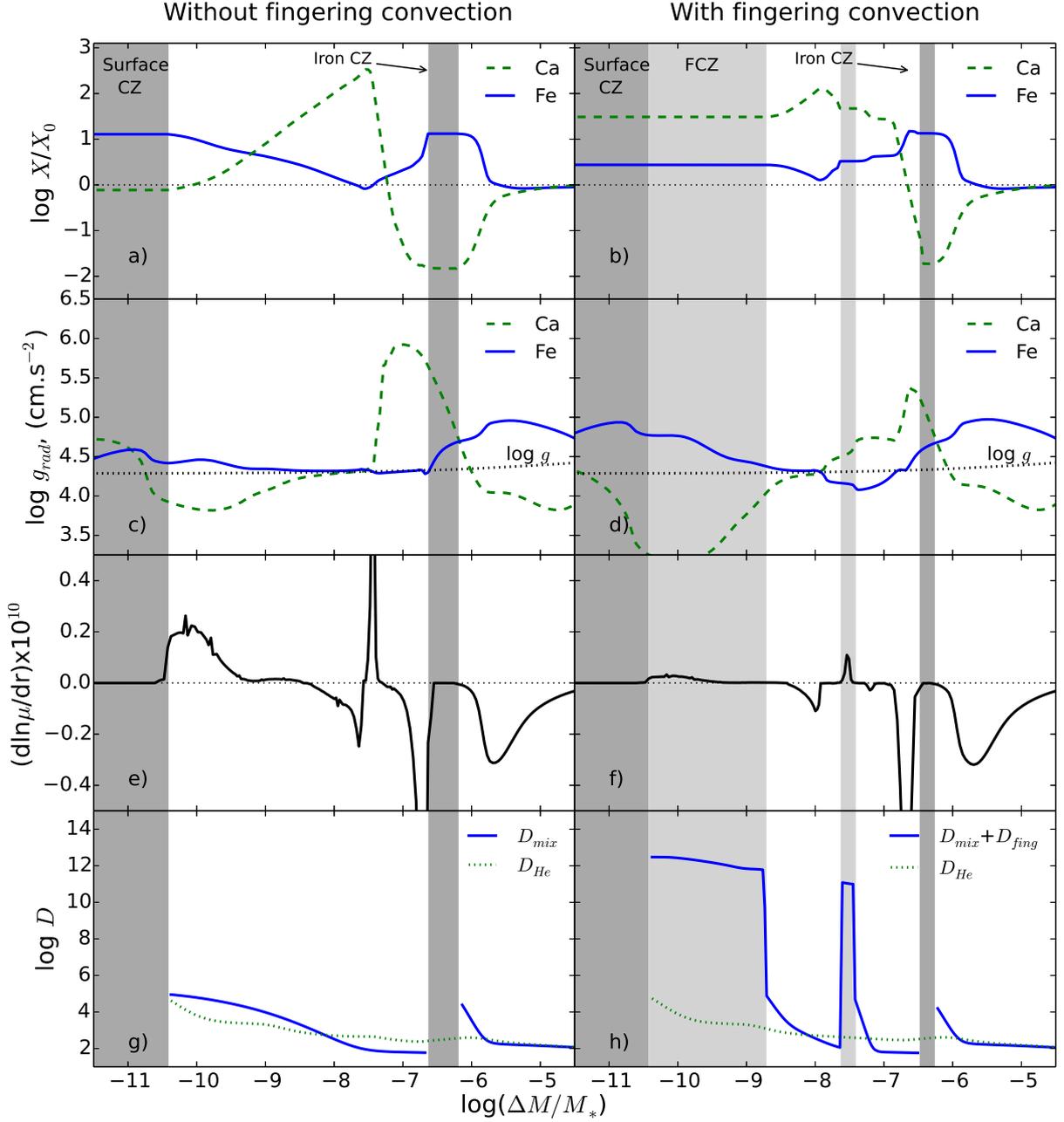}
\caption{ Profiles of important physical quantities as a function of $\log(\Delta M/M_*)$ in two 1.7 $M_{\odot}$ models at 100~Myrs with (right panel) and without (left panels) computation of fingering convection. Dynamical(CZ) and fingering (FCZ) convective zones are represented by dark and light grey regions, respectively. Panels \textit{a} and \textit{b} show calcium (green dashed lines) and iron (blue solid lines) abundances compare to their initial value. Panels \textit{c} and \textit{d} show calcium (green dashed lines) and iron (blue solid lines) radiative accelerations. The black dotted line represents the gravity. Panels \textit{e} and \textit{f} show the $\ln\mu$-gradient with radius (black solid lines), panels \textit{g} and \textit{h} show the helium diffusion coefficient (green dotted lines) and effective mixing coefficient (blue solid lines).}
\label{fig1}
\end{center}
\end{figure*}

\subsection{Treatment of fingering convection}

Fingering (thermohaline) convection is a well-known process in oceanography. This instability occurs when hot salt water comes upon cool less salt water. It is indeed at the origin of the global circulation in the Earth oceans, called "thermohaline circulation". In stars, a similar instability occurs every time heavy matter comes upon lighter matter in the presence of a stable temperature gradient. This may happen in the case of accretion of planetary matter \citep{vauclair04, garaud11, deal13} or accretion of matter from a companion onto the star \citep{stancliffe08,thompson08}. Fingering convection was also invoked in the case of a local $\mu$-decrease due to nuclear reactions as in Red Giants \citep{charbonnel07}, although the effect is too small to account for the observations (e.g. \citealt{wachlin14}). We are interested in the case of a local heavy element accumulation due to radiative accelerations that lead to an increase of $\mu$.

Fingering convection is characterised by the so-called density ratio $R_0$, which is the ratio between thermal and $\mu$-gradients, i.e. 
\begin{equation}
R_0=\frac{\delta}{\phi}\frac{\nabla - \nabla_{ad}}{\nabla_{\mu}}. 
\end{equation} 
This instability can only develop if the thermal diffusivity is larger than the molecular diffusivity. In this case, the heavy blobs of fluid fall inside the star and keep falling because they thermalise faster than the heavy elements diffuse outside the star. The blobs stop falling when the thermal and diffusion fluxes equilibrate. In any case, fingering convection cannot occur if $R_0$ is larger than the ratio of the thermal to the molecular diffusivities. This leads to
\begin{equation}
1<R_0<\frac{1}{\tau} 
,\end{equation}  
where $\tau$ is the inverse Lewis number, ratio of molecular, and thermal diffusivity.
For values of $R_0<1$ the region is dynamically convective (Ledoux criteria), and for values of $R_0>1/\tau$ the region is stable.

The first treatments of fingering convection in stars were purely analytical \citep{ulrich72,kippenhahn80}. Recently, two-dimensional (2D) and three-dimensional (3D) numerical simulations were performed, all converging on the result that the \citet{ulrich72} value was strongly overestimated \citep{denissenkov10,traxler11}. 

The recent 3D simulations by \citet{brown13} (hereafter BGS),  including the evolution of the fingers with time, yield coefficients slightly larger than the previous coefficients. These authors deduce a prescription for 1D models from their simulations with an effective fingering diffusion coefficient given by
\begin{equation}
D_{fing}=Nu_{\mu}\kappa_{\mu},\end{equation}  
where $Nu_{\mu}$ is the Nusselt number and $\kappa_{\mu}$ is the molecular diffusivity.

This work was performed using the BGS prescription for the computation of fingering convection, which represents a real improvement compared to the previous treatments (see also \citealt{zemskova14}).

\section{Results}

\subsection{Computations of fingering convection in 1.7 $M_{\odot}$ models}

\begin{figure}
\begin{center}
\includegraphics[width=0.45\textwidth]{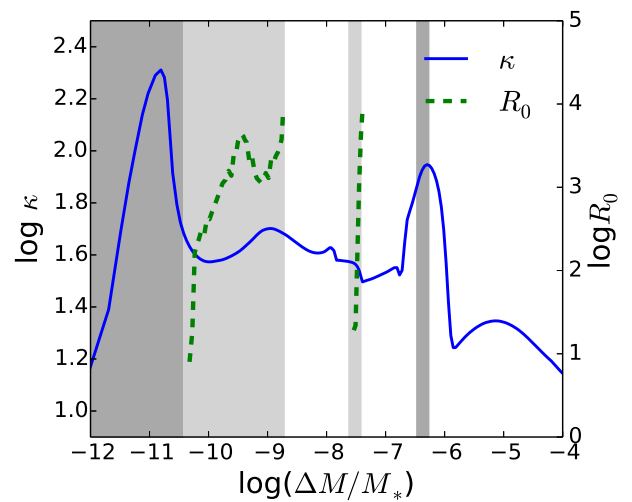}
\caption{Rosseland opacity profile in a 1.7~$M_{\odot}$ model including the complete effects of atomic diffusion and fingering convection (blue solid line). The values of the $R_0$ parameter are also shown inside the fingering zones (green dashed lines). In the same zones, the Lewis number is of order $10^{10}$.}
\label{fig2}
\end{center}
\end{figure}

The case of 1.7~$M_{\odot}$ models were studied by \cite{theado09} in a preliminary way. They showed that this instability could occur at the bottom of the iron convective zone induced by atomic diffusion, which appears at $T\approx 200~000~K$. They discussed how fingering convection reduces the accumulation of iron in this region without stopping it.

We compute fingering convection over the entire star. We show how the accumulations of elements other than iron also play a role in the overall fingering mixing.  As a first step, we add no other extra-mixing processes such as rotation or internal wave, as the aim of these computations is to test the specific effect of fingering convection on the stellar structure.

Fig. \ref{fig1} compares the results obtained in 1.7 $M_{\odot}$ models without and with fingering convection, after 100~Myrs. 
Panels \textit{a} and \textit{b} show the abundance profiles of Ca and Fe, panels \textit{c} and \textit{d} show the radiative accelerations on these elements, panels \textit{e} and \textit{f} show the local $\mu$-gradients, and panels \textit{g} and \textit{h} show the various effective diffusion coefficients implied in the computations including the helium atomic diffusion coefficient.

In both cases the opacity increase, which is induced by the iron accumulation, induces a dynamical convective zone at $\log(\Delta M/M_*)\approx-6.3$. This effect is due to the fact that iron accumulates at the place where it is the main contributor of the overall opacity (see Fig. \ref{fig2}). When fingering convection is taken into account, it reduces the iron accumulation, leading to a narrower convective zone. 

We can see on the right panels that two regions of fingering instabilities (hereafter FCZ, for fingering convective zones) appear between the surface dynamical convective zone and the iron convective zone. These regions are unstable owing to $\mu$-gradient inversions. In these zones, the parameter $R_0$ is always larger than 1 and smaller than the Lewis number, which has values of order $10^{10}$ (see Fig. 2). The first zone is due to the heavy elements accumulation inside the surface convective zone. The second zone is induced by the local calcium accumulation, which occurs around $\log(\Delta M/M_*)\approx-7.5$, as the result of the rapid upwards decrease of the radiative acceleration on this element. 

The first FCZ mixes the stellar gas down to $\log(\Delta M/M_*)\approx-8.8$ and reduces the iron surface abundance by a factor 3. Meanwhile, the surface calcium abundance is increased by a factor 40 (blue solid and green dashed lines of panel \textit{b}, respectively) as a result of this mixing. An interesting point is that, while the fingering convection mixing reduces the iron concentration, it leads to an increase of g$_{rad}$(Fe) (blue solid line of \textit{d} panel) because radiative accelerations are smaller when the abundance increases. This maintains the existence of iron accumulation and induced fingering convection in this region. 

The second FCZ was a surprise because we did not anticipate that the accumulation of an element other than iron or nickel could produce an unstable $\mu$-gradient in this region. This narrow FCZ mixes the elements in a small region and moves with time inside the star (see section 3.2). This effect leads to an increase of the abundances of iron and calcium between $-7.5<\log(\Delta M/M_*)<-6.8$ (blue solid and green dashed lines of panel \textit{b}, respectively). Because of the increase of the calcium abundance in this region, the radiative acceleration decreases by a factor 4 (green dashed line of \textit{d} panel).

\subsection{The evolution of hydrodynamical instabilities with stellar age}  

\begin{figure}
\begin{center}
\includegraphics[width=0.5\textwidth]{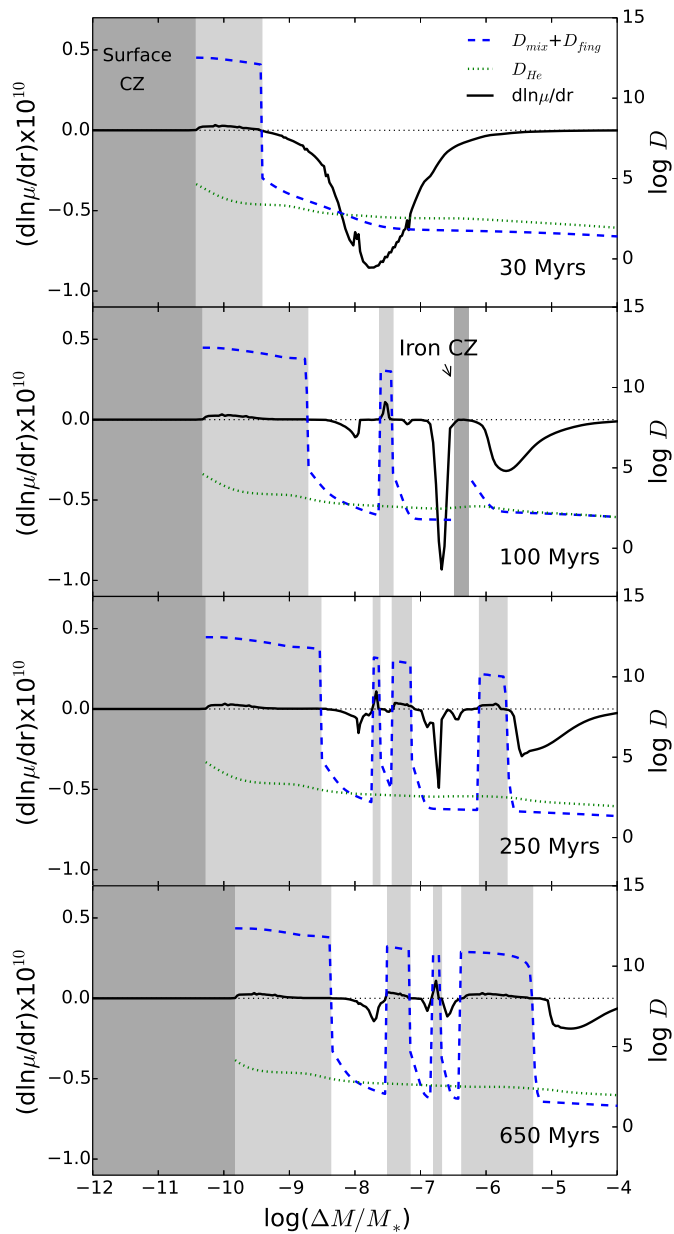}
\caption{Profiles of the $\mu$-gradient as a function of $\log(\Delta M/M_*)$ (black solid lines), helium diffusion coefficient (green dotted lines) and effective mixing coefficient (blue dashed lines) at four ages. Dynamical and fingering convective zones are represented by dark and light grey regions, respectively.}
\label{fig3}
\end{center}
\end{figure}

\begin{figure}
\begin{center}
\includegraphics[width=0.49\textwidth]{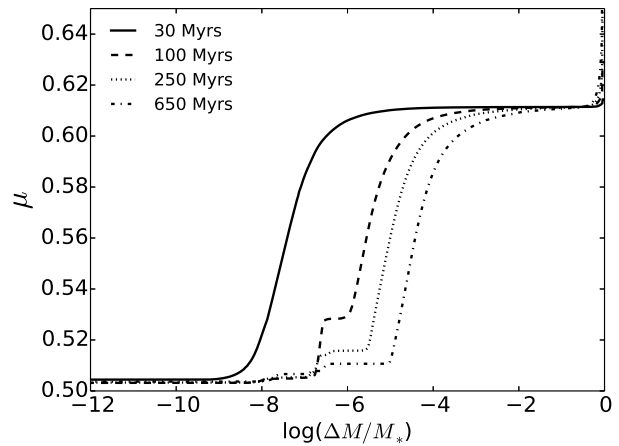}
\caption{$\mu$-profiles of a 1.7 $M_{\odot}$ with fingering convection at four ages.}
\label{fig4}
\end{center}
\end{figure}

Fig. \ref{fig3} shows the profiles as a function of $\log(\Delta M/M_*)$ of the $\mu$-gradient (black solid lines), the helium diffusion coefficient (green dotted lines), and the global effective mixing coefficient $D _{mix}+D _{fing}$ (blue dashed lines) in the 1.7 $M_{\odot}$ models, including fingering convection at 30, 100, 250, and 650~Myrs (from upper to lower panel). The FCZ at the bottom of the surface convective zone is maintained during all the main sequence and mixes the star deeper as the star evolves. The FCZ triggered by the accumulation of calcium still appears after 100~Myrs but moves with time. This FCZ slightly varies in size and position as the star evolves and is also split at some times. This explains the behaviour of the global effective mixing diffusion coefficient around $\log(\Delta M/M_*)\approx -7.6 $ at 250~Myrs and around $\log(\Delta M/M_*)\approx -6.8 $ at 650~Myrs, and it occurs all over the main sequence.
 
The reason why there is no FCZ below the iron convective zone at 100~Myrs is because of the helium stable gradient that prevents this instability at this age. As shown in Fig.\ref{fig4}, the downwards helium diffusion builds up a stable $\mu$-gradient that sinks inside the star with time. The heavy elements accumulation can overcome this helium-induced $\mu$-gradient and lead to an unstable  $\mu$-gradient later on. After roughly 250~Myrs, fingering convection occurs and reduces the concentration of iron in this region. The opacity is reduced and the dynamical convective zone disappears, but the layers are still mixed by fingering convection. This mixed region then extends with time as the helium gradient moves downwards. 

\subsection{Influence of the stellar mass}  

We computed 1.5 and 1.9 $M_{\odot}$ models with the same approach as the 1.7 $M_{\odot}$ models. The main differences between these three models are related to the depth of the surface convective zone, which narrows for increasing stellar masses, and to the intensity of the radiative accelerations, which increase with masses. For these reasons, the local accumulations of heavy elements are less important in a 1.5 $M_{\odot}$ star than in a 1.9 $M_{\odot}$. 

\begin{figure}
\begin{center}
\includegraphics[width=0.49\textwidth]{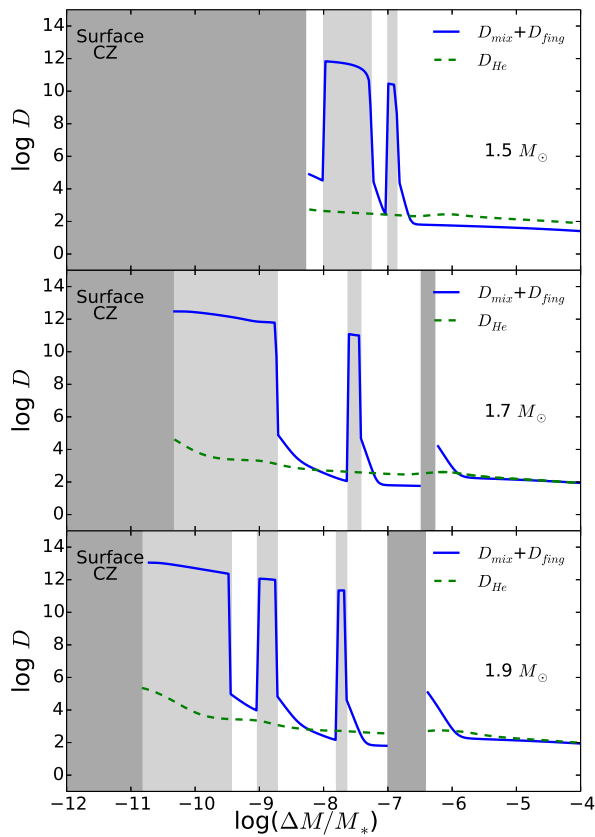}
\caption{Global effective mixing diffusion coefficient and helium diffusion coefficient profiles at 100~Myrs three masses with fingering convection. }
\label{fig5}
\end{center}
\end{figure}

Fig. \ref{fig5} shows that the effective fingering diffusion coefficient of 1.5, 1.7, and 1.9 $M_{\odot}$ models at 100~Myrs. FCZs are present in all models. Accumulations of iron and calcium are more likely to happen in more massive stars because of stronger radiative accelerations. This produces stronger unstable $\mu$-gradients and triggers a more efficient fingering convection mixing.

Therefore, it is easy to imagine that this kind of mixing can appear in more massive stars if extra mixing process are not strong enough to prevent the accumulations of elements, such as iron or calcium, by atomic diffusion. Further studies on more massive stars should be done to characterise the impact of fingering convection in these cases.

\subsection{Results on surface abundances}

\begin{figure}
\begin{center}
\includegraphics[width=0.51\textwidth]{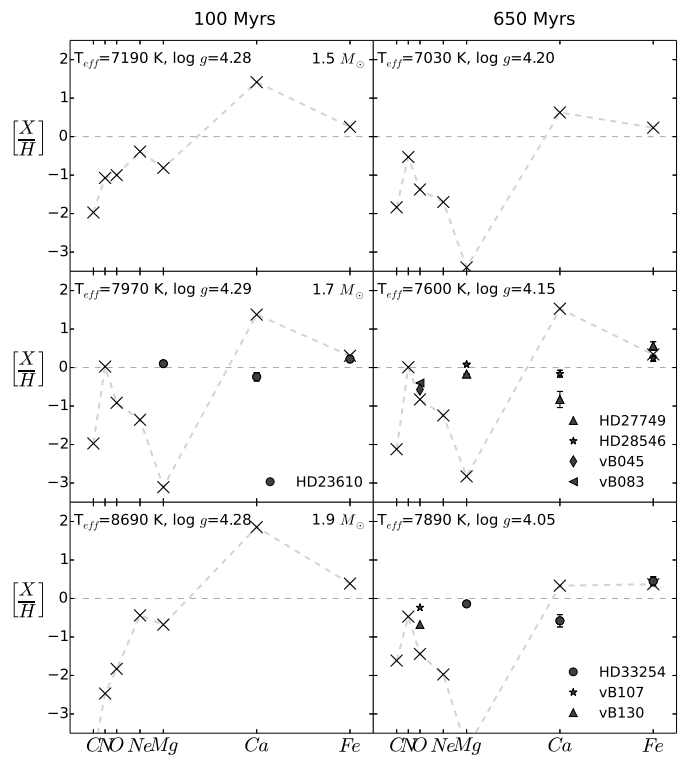}
\caption{Surface abundances at 100~Myrs (left panel) and 650~Myrs (right panel) predicted by 1.5 $M_{\odot}$ (upper panels), 1.7 $M_{\odot}$ (middle panels), and 1.9 $M_{\odot}$ (lower panels) models with fingering convection represented by black crosses and light grey dashed lines. The abundances obtained from observations are represented by black symbols.}
\label{ab}
\end{center}
\end{figure}

The accumulation of heavy elements inside the surface convective zone induces a FCZ that mixes the chemical elements deeper inside the star. This produces visible effects in the surface chemical composition that can be compared to the abundance determinations obtained from observations. Fig. \ref{ab} shows surface abundances predicted by 1.5 $M_{\odot}$ (upper panel), 1.7 $M_{\odot}$ (middle panel), and 1.9 $M_{\odot}$ (lower panel) models including fingering convection. These are compared to those observed in Hyades and Pleiades stars with similar effective temperature and surface gravity when available. Observations of HD 23610 (dots), HD 27749 (triangles), and HD 28546 (stars) from \cite{huibonhoa98}, and vB 045 (rhombus) and vB 083 (horizontal triangle) from \cite{takeda97} are used for comparison with the 1.7 $M_{\odot}$ (middle panels). Observations of HD 33254 (dots) from \cite{huibonhoa98} and vB 107 (star) and vB 130 (triangle) from \cite{takeda97} are used for comparison with the 1.9 $M_{\odot}$ (lower panels). Even if we do not have enough observations at 100 and 650 Myrs, we represent the results of the computations for 1.5 $M_{\odot}$ (upper panels) and 1.9 $M_{\odot}$ (lower left panel).

\begin{figure}
\includegraphics[width=0.51\textwidth]{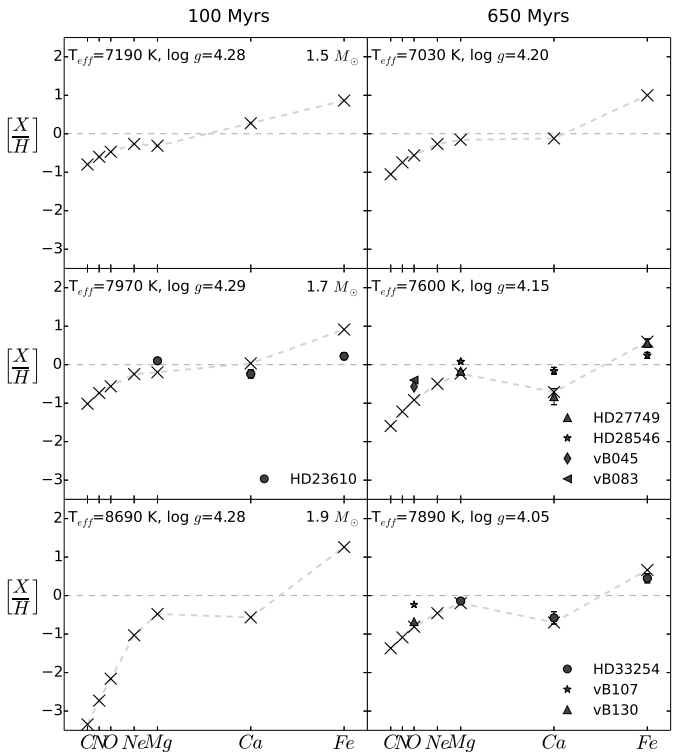}
\caption{Same as Fig. \ref{ab} with fingering convection and connections between FCZs and dynamical convective zones. }
\label{ab-connect}
\end{figure}

We can see that, in spite of the extra mixing induced by fingering convection, there is no concordance between the abundances determined from observations and those predicted by our models. In these models however, no overshooting has been introduced between the mixed zones induced by fingering convection. If we assume that these regions are connected by overshooting, the results are modified. We computed 1.5, 1.7, and 1.9 $M_{\odot}$ models in which we assumed connections between the surface dynamical convective zone and the deepest mixed zone (iron or fingering convective zone). In this case, the surface abundances become strikingly similar to the abundances determined from observations (Fig. \ref{ab-connect}). This confirms the results of \cite{richard01} showing that a mixing down to $\log(\Delta M/M_*)\approx-6.5$ is needed to reproduce the abundances of Am stars.

\section{Discussion and conclusions}
\label{discussion}

In this paper, we confirm that atomic diffusion, a microscopic process acting on individual atoms, has important macroscopic consequences on the stellar structure, hydrodynamical instabilities at work inside the stars, and on the resulting surface abundances.

Fingering convection, which represents the only macroscopic process directly induced by the local accumulation of heavy elements in specific stellar layers, cannot be forgotten in the computations of the evolution of A-type stars. Contrary to other processes like rotation-induced mixing or mass loss, it is not arbitrarily added to explain the observations. It is an unavoidable consequence of the diffusion-induced inverse $\mu$-gradient.

We show that if this process is the only hydrodynamical instability taken into account in the computations, the stellar gas below the outer convective zones is locally mixed in several layers, which move with time during stellar evolution. If we only take these mixed zones into account in the computations, we do not find surface abundances that fit correctly the observations of Am stars. The fit becomes much better if we assume that all these zones are connected by overshooting.

In real stars, rotational mixing is clearly important to account for the statistical observations. Chemically peculiar stars, such as Am stars, are known to be slow rotators, whereas stars with normal abundances, such as $\delta$-Scuti stars, rotate more rapidly. 

The effect of rotational-induced mixing is primarily to reduce and eventually suppress the abundance variations induced by atomic diffusion in rapidly rotating stars. In slow rotators, the influence of meridional circulation on the overall process should also be tested.

Mass loss could also have an important effect on abundance stratification in stars \citep{vick10,michaud11}. It could modify the depth or appearance of fingering convective zones.

The next step will be to discuss the interaction between rotation-induced mixing and fingering convection to model these stars in a more complete and realistic way. This is not trivial as the external hydrodynamical instability may modify the formation of the fingers, which represent the convective cells of this type of double-diffusive convection. At the present time, we can only infer that in slow rotators the meridional circulation may help connect the fingering zones, which is needed to explain quantitatively the abundances derived from the observations.  

\begin{acknowledgements}

\end{acknowledgements}


\bibliographystyle{aa.bst} 
\bibliography{biblio} 

\end{document}